\documentclass[journal]{IEEEtran}
\IEEEoverridecommandlockouts

\usepackage{cite}
\usepackage{amsmath,amssymb,amsfonts}
\usepackage{algorithmic}
\usepackage{graphicx}
\usepackage{textcomp}
\usepackage{xcolor}
\usepackage{tabularx}
\usepackage{booktabs}
\usepackage[colorlinks,linkcolor=red,anchorcolor=red,citecolor=blue]{hyperref}
\usepackage[lined,boxed,ruled]{algorithm2e}

\usepackage{tikz}
\usetikzlibrary{shapes.geometric, arrows.meta, positioning, fit, backgrounds, calc}
\usetikzlibrary{arrows.meta, positioning, shapes.geometric, shadows, patterns}

\usetikzlibrary{shapes.geometric, fit, backgrounds, arrows.meta}

\usepackage{pgfplots}
\pgfplotsset{compat=1.18}

\graphicspath{{./figs/}}

\def\BibTeX{{\rm B\kern-.05em{\sc i\kern-.025em b}\kern-.08em
    T\kern-.1667em\lower.7ex\hbox{E}\kern-.125emX}}
\begin{document}

\title{Decoupling Speculation from Merit: The Identity-Bound Asset Integrity Model (IBAIM) for Sustainable Web3 Gaming}

\author{\IEEEauthorblockN{Jinliang Xu$^{*}$}\\
\IEEEauthorblockA{\textit{China Academy of Information and Communications Technology}, Beijing, China \\
xujinliang@caict.ac.cn}\\
}
\IEEEaftertitletext{\vspace{-1\baselineskip}}
\maketitle

\begin{abstract}
The rapid collapse of decentralized game economies, often characterized by the \textit{death spiral,} remains the most formidable barrier to the mass adoption of Web3 gaming. This paper proposes that the sustainability of an open game economy is predicated on three necessary and sufficient conditions: Anti-Sybil Resilience, Anti-Capital Dominance, and Anti-Inflationary Saturation.
The first section establishes a theoretical proof of these conditions, arguing that the absence of any single dimension leads to systemic failure. The second section explores the dialectical relationship between these dimensions, illustrating how unchecked automation and capital-driven monopolies accelerate asset hyperinflation.
In the third section, we introduce the Identity-Bound Asset Integrity Model (IBAIM) as a comprehensive technical solution. 
IBAIM utilizes Zero-Knowledge (ZK) biometric hashing and Account Abstraction (AA) to anchor asset utility to unique human identities through a privacy-preserving and regulatory-compliant architecture. By exogenizing biometric verification to trusted local environments and utilizing Zero-Knowledge Proofs of Identity (zk-PoI), the model ensures absolute user privacy. Furthermore, by implementing an Asymmetric Utility Decay (AUD) engine—whereby assets suffer a vertical 50\% utility cliff upon secondary transfer—and an entropy-driven thermodynamic degradation mechanism., the model successfully decouples financial speculation from in-game merit.
Finally, we apply this framework to analyze prominent historical failures in the GameFi sector, demonstrating that their collapse was an inevitable consequence of violating these core economic constraints. Our findings suggest that trading a degree of asset liquidity for system integrity is the only viable path toward long-term economic viability in decentralized virtual worlds.

\end{abstract}

\section{Introduction}
The emergence of Web3 technologies has catalyzed a paradigm shift in digital entertainment, giving rise to the Open Game Economy (OGE). Unlike traditional closed game ecosystems, Web3 games leverage blockchain technology and non-fungible tokens (NFTs) to grant players true ownership of digital assets and the freedom to trade them on secondary markets. Early models, particularly \textit{Play-to-Earn} (P2E), promised to blur the lines between virtual leisure and real-world financial gain. However, this high degree of financialization has exposed severe systemic vulnerabilities. Exposing in-game economies to external speculative capital and frictionless secondary markets has repeatedly resulted in a catastrophic \textit{death spiral} across the decentralized gaming sector: an unchecked oversupply of tokens coupled with a collapse in demand, leading to hyperinflation and liquidity exhaustion. 

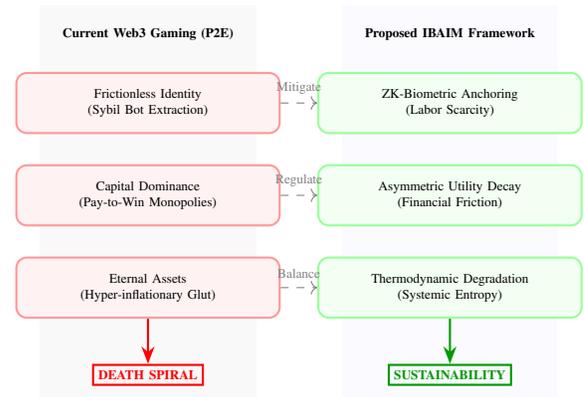
\begin{figure}[htbp]
\centering
\begin{tikzpicture}[
    node distance=1.2cm,
    font=\tiny,
    box/.style={rectangle, draw, thick, rounded corners, align=center, minimum width=3.5cm, minimum height=0.8cm},
    problem/.style={box, fill=red!5, draw=red!40},
    solution/.style={box, fill=green!5, draw=green!40},
    arrow/.style={-Stealth, thick},
    label/.style={font=\tiny\bfseries}
]

    \node [label] (left_title) {Current Web3 Gaming (P2E)};
    \node [problem, below=0.3cm of left_title] (p1) {Frictionless Identity\\(Sybil Bot Extraction)};
    \node [problem, below=0.4cm of p1] (p2) {Capital Dominance\\(Pay-to-Win Monopolies)};
    \node [problem, below=0.4cm of p2] (p3) {Eternal Assets\\(Hyper-inflationary Glut)};
    
    \node [draw, red, thick, inner sep=2pt, below=0.6cm of p3] (spiral) {\textbf{DEATH SPIRAL}};
    \draw [arrow, red] (p3) -- (spiral);

    \node [label, right=1.5cm of left_title] (right_title) {Proposed IBAIM Framework};
    \node [solution, below=0.3cm of right_title] (s1) {ZK-Biometric Anchoring\\(Labor Scarcity)};
    \node [solution, below=0.4cm of s1] (s2) {Asymmetric Utility Decay\\(Financial Friction)};
    \node [solution, below=0.4cm of s2] (s3) {Thermodynamic Degradation\\(Systemic Entropy)};

    \node [draw, green!60!black, thick, inner sep=2pt, below=0.6cm of s3] (stable) {\textbf{SUSTAINABILITY}};
    \draw [arrow, green!60!black] (s3) -- (stable);

    % --- 关联箭头 ---
    \draw [dashed, gray, ->] (p1.east) -- node[above, sloped] {Mitigate} (s1.west);
    \draw [dashed, gray, ->] (p2.east) -- node[above, sloped] {Regulate} (s2.west);
    \draw [dashed, gray, ->] (p3.east) -- node[above, sloped] {Balance} (s3.west);

    % 背景框 (可选，增强区分度)
    \begin{scope}[on background layer]
        \node [fill=gray!5, fit=(left_title) (spiral), inner sep=5pt] {};
        \node [fill=blue!2, fit=(right_title) (stable), inner sep=5pt] {};
    \end{scope}

\end{tikzpicture}
\caption{The paradigm shift from speculative \textit{Ponzi-nomics} to IBAIM. The framework addresses the core misalignment in decentralized economies by introducing identity scarcity, capital friction, and digital entropy.}
\label{fig:intro_paradigm}
\end{figure}

The root of this systemic failure lies in a fundamental misalignment of economic design across three dimensions. First is the absence of stringent identity verification \cite{ohlhaver2022decentralized,fiege1987zero}. When the cost of creating a new digital identity is frictionless, the ecosystem becomes highly susceptible to Sybil attacks, allowing automated scripts and bot farms to infinitely extract the reward pool. Second is the unrestrained dominance of capital. When financial wealth can seamlessly and infinitely substitute for actual gameplay effort, the ecosystem devolves into a monopolistic \textit{Pay-to-Win} structure. This alienates the core labor class—genuine players—whose exodus inevitably destroys the fundamental demand and utility of the assets. Third is the paradox of \textit{eternal assets.} Unlike physical goods that naturally decay, decentralized digital assets are mathematically immortal. Without robust mechanisms for digital entropy and asset destruction, the continuous minting of rewards inevitably outpaces demand, triggering inescapable hyperinflation. The visual representation of this paradigm shift, from current speculative models to the proposed sustainable framework, is illustrated in Fig. \ref{fig:intro_paradigm}.

To address this seemingly cursed problem and achieve long-term economic sustainability, this paper proposes that an Open Game Economy must satisfy three necessary and sufficient conditions: Anti-Sybil Resilience, Anti-Capital Dominance, and Anti-Inflationary Saturation. The absence of any single dimension guarantees the eventual collapse of the entire economic structure.

Grounded in this theoretical framework, this paper introduces the Identity-Bound Asset Integrity Model (IBAIM), a comprehensive economic and technical architecture designed to decouple financial speculation from in-game merit. IBAIM employs Zero-Knowledge (ZK) biometric hashing and Account Abstraction to firmly anchor digital asset production and utility to verified, unique human labor. To counteract capital monopolies, the model introduces an \textit{Asymmetric Utility Decay} mechanism, which imposes a strict performance penalty on assets transferred through secondary markets. Furthermore, it enforces a dual-factor thermodynamic degradation system where assets inevitably decay through time and usage, requiring continuous human labor for repair and maintenance. Ultimately, this research argues that transitioning from speculative \textit{Ponzi-nomics} to a sustainable, labor-centric value system requires a deliberate trade-off: sacrificing a degree of absolute asset liquidity is the only viable path to preserving the systemic integrity and longevity of decentralized virtual worlds.

\section{Necessary and Sufficient Conditions for Economic Sustainability}

The emergence of decentralized technologies has birthed the \textit{Open Game Economy} (OGE), a digital ecosystem characterized by permissionless asset ownership and secondary market liquidity. However, the majority of extant OGEs have succumbed to the \textit{Death Spiral}-a terminal state of hyperinflation and liquidity exhaustion. To address this, we must rigorously define the theoretical boundaries of economic sustainability.

\subsection{Defining Sustainability in Decentralized Systems}

Economic sustainability in an OGE is defined as the system's ability to maintain a non-zero marginal value of its core assets ($V_a > 0$) and a stable circulation of utility without reliance on exogenous capital injections. Formally, a system is sustainable if for any time $t$, the net utility produced by participants equals or exceeds the utility extracted by profit-seeking agents, adjusted for systemic entropy.

This equilibrium fundamentally relies on the principle of \textbf{Incentive Compatibility} \cite{xu2018reward}, where a structured framework of \textbf{Reward and Punishment} is employed to ensure that the rational self-interest of individual participants aligns with the long-term collective stability of the ecosystem.

\subsection{The Necessary Conditions: A Proof by Omission}

We posit that three fundamental dimensions constitute the necessary conditions for OGE sustainability. The necessity of these conditions is demonstrated by the systemic failure observed when any single dimension is absent.

\paragraph{Anti-Sybil Resilience ($\mathcal{C}_{Sybil}$)}
In an open economy, if the marginal cost of identity creation $\epsilon$ is lower than the expected return on automated labor $R(l)$, the system will be overwhelmed by Sybil attacks. 
\begin{equation}
    \text{Necessity: } \text{If } \mathcal{C}_{Sybil} \to 0, \text{ then } \text{Supply} \to \infty, \text{ forcing } V_a \to 0.
\end{equation}
Without a robust identity-anchoring mechanism, the reward pool is exhausted by non-human actors, leading to the \textit{Tragedy of the Digital Commons.}

\paragraph{Anti-Capital Dominance ($\mathcal{C}_{Capital}$)}
Sustainability requires a separation between financial investment and meritocratic progression. If capital can directly and infinitely substitute for gameplay effort (labor), the economy shifts from a \textit{Play-and-Earn} model to a \textit{Pay-to-Win} (P2W) monopoly.
\begin{equation}
    \text{Necessity: } \text{If } \frac{\partial \text{Utility}}{\partial \text{Capital}} \gg \frac{\partial \text{Utility}}{\partial \text{Labor}}, \text{ then } \text{Retention} \to 0.
\end{equation}
The exodus of the \textit{labor class} (players) destroys the utility demand, rendering the assets held by the \textit{capital class} worthless.

\paragraph{Anti-Inflationary Saturation ($\mathcal{C}_{Inflation}$)}
Decentralized assets are often permanent, while utility is ephemeral. Without a mechanism to increase systemic entropy (asset destruction), the total supply $S$ will eventually exceed the saturation point of demand $D$.
\begin{equation}
    \text{Necessity: } \text{If } \oint \text{Minting} > \oint \text{Burning}, \text{ then } \Delta V_a < 0.
\end{equation}
Continuous supply growth without an equivalent \textit{sink} necessitates the inevitable devaluation of the unit price.

\subsection{The Sufficient Condition: The ICR Closed-Loop Model}

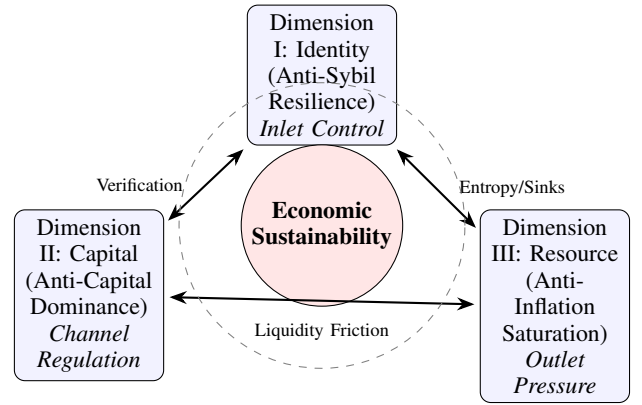
\begin{figure}[htbp]
\centering
\begin{tikzpicture}[
    node distance=1.2cm,
    block/.style={rectangle, draw, fill=blue!5, text width=1.75cm, align=center, rounded corners, minimum height=0.75cm, font=\small},
    arrow/.style={thick, <->, >=Stealth, shorten >=2pt, shorten <=2pt},
    centerblock/.style={circle, draw, fill=red!10, align=center, font=\small\bfseries, inner sep=2pt}
]

    \node [block] (identity) {Dimension I: Identity\\(Anti-Sybil Resilience)\\ \textnormal{\textit{Inlet Control}}};
    
    \node [block, below left=of identity, xshift=-0.25cm] (capital) {Dimension II: Capital\\(Anti-Capital Dominance)\\ \textnormal{\textit{Channel Regulation}}};
    
    \node [block, below right=of identity, xshift=0.25cm] (resource) {Dimension III: Resource\\(Anti-Inflation Saturation)\\ \textnormal{\textit{Outlet Pressure}}};

    \node [centerblock] (center) at (barycentric cs:identity=1,capital=1,resource=1) {Economic\\Sustainability};

    \draw [arrow] (identity) -- node[left, font=\scriptsize, xshift=-5pt] {Verification} (capital);
    \draw [arrow] (capital) -- node[below, font=\scriptsize, yshift=-5pt] {Liquidity Friction} (resource);
    \draw [arrow] (resource) -- node[right, font=\scriptsize, xshift=5pt] {Entropy/Sinks} (identity);

    \draw [dashed, gray] (center) circle (1.9cm);

\end{tikzpicture}
\caption{The ICR Framework: Necessary and Sufficient Conditions for Economic Sustainability.}
\label{fig:icr_model}
\end{figure}

While the aforementioned conditions are individually necessary, we argue they are collectively sufficient to ensure an internal equilibrium. We propose the Identity-Capital-Resource (ICR) framework as a closed-loop model, as depicted in Fig. \ref{fig:icr_model}. 

Let $\mathcal{S}$ be the sustainability of the economy. We define $\mathcal{S}$ as a function of the inlet control (Identity), the channel regulation (Capital), and the outlet pressure (Resource/Inflation):
\begin{equation}
    \mathcal{S} \iff \mathcal{C}_{Sybil} \land \mathcal{C}_{Capital} \land \mathcal{C}_{Inflation}
\end{equation}

The sufficiency is grounded in the total coverage of the asset lifecycle:
\begin{enumerate}
    \item \textbf{Inlet Control (Identity):} Regulates the rate of entry and validates the authenticity of production.
    \item \textbf{Channel Regulation (Capital):} Imposes \textit{friction} on the financialization of utility, ensuring that the velocity of capital does not destabilize the game’s meritocratic core.
    \item \textbf{Outlet Pressure (Resource):} Implements a \textit{sink} through dynamic degradation, ensuring that for every unit of value produced, a proportional unit is consumed or depreciated.
\end{enumerate}

By satisfying these three dimensions, the OGE moves from a linear extractive model to a circular regenerative model. This tri-dimensional equilibrium ensures that the economy can withstand both internal fluctuations and external speculative shocks, providing a theoretical foundation for the long-term survival of virtual digital societies.

\section{The Tri-Dimensional Framework: Definitions and Dialectics}

To construct a robust economic defense, we must move beyond the superficial understanding of game mechanics and analyze the underlying forces of Identity, Capital, and Resource Inflation. This section provides a granular definition of each dimension and explores their interdependencies within a decentralized ecosystem.

\subsection{Dimension I: Anti-Sybil and the Verification of Human Labor}

In an OGE, the primary source of value is often the time and effort invested by participants, termed here as \textit{Proof-of-Play-Effort} (PoPE). The fundamental threat to this value is the decoupling of the biological human from the digital agent \cite{fiege1987zero}.

Anti-Sybil resilience ($\mathcal{C}_{Sybil}$) is not merely a technical barrier; it is the gatekeeper of labor scarcity. When identity creation is frictionless, automated scripts (bots) can simulate human labor at scale. We define the \textit{Identity-Labor Coefficient} $\lambda$ as:
\begin{equation}
    \lambda = \frac{\text{Verified Human Actors}}{\text{Total Active Accounts}}
\end{equation}
A sustainable system must maintain $\lambda \to 1$. If $\lambda$ drops, the marginal utility of human effort is diluted, leading to the collapse of the meritocratic reward structure.

\subsection{Dimension II: Anti-Capital Dominance and Utility Friction}

Capital dominance ($\mathcal{C}_{Capital}$) occurs when the liquid wealth of a participant allows them to bypass the intended progression curve of the game. In Web3, this is exacerbated by high-velocity secondary markets.

The core objective of this dimension is to introduce \textit{Utility Friction}. We propose that the utility $U$ of an asset should be a function of both its inherent properties $P$ and its \textit{Attribution Status} $A$.
\begin{equation}
    U = f(P, A)
\end{equation}
Where $A=1$ for the original creator (First-hand) and $A < 1$ for secondary holders (Second-hand). Instead of a mere penalty, this mechanism acts as an 'attunement' requirement or {\bf stranger tax}; while capital can acquire the asset and immediately gain 50\% of its utility—which is still highly beneficial for players to jumpstart their progress—it cannot fully harness the original creator's labor-earned potency, thus preserving the prestige of genuine labor. This friction prevents \textit{Whale-monopolization} where a small percentage of wealthy actors own 100\% of the server's effective power.

\subsection{Dimension III: Anti-Inflation and Digital Entropy}

The \textit{Eternal Asset} paradox is the primary driver of OGE inflation. In traditional economies, physical goods naturally decay. In digital economies, assets are mathematically immortal, leading to a monotonic increase in total supply $S_{total}$.

Anti-inflationary saturation ($\mathcal{C}_{Inflation}$) requires the introduction of \textit{Digital Entropy} $\delta$. Total supply at time $t$ must follow:
\begin{equation}
    S(t) = \int_0^t (M(\tau) - \delta(S, U, \tau)) d\tau
\end{equation}
Where $M(\tau)$ is the minting rate and $\delta$ is the degradation rate. Sustainability is achieved when $\delta$ scales with usage and time, effectively creating a \textit{Resource Sink} that necessitates continuous production and replacement, thereby maintaining demand for newly minted labor-based assets.

\subsection{The Interdependency: A Feedback Loop Analysis}

The three dimensions do not exist in isolation; they form a reflexive loop. Their relationship can be characterized by the following synergistic effects:

\begin{itemize}
    \item \textbf{Identity-Inflation Linkage:} If $\mathcal{C}_{Sybil}$ is weak, the minting rate $M(\tau)$ becomes uncontrollable, rendering any fixed degradation rate $\delta$ insufficient to prevent devaluation.
    \item \textbf{Capital-Identity Linkage:} If $\mathcal{C}_{Capital}$ is not managed, wealthy actors will seek to \textit{buy} verified identities (account trading), effectively bypassing the Anti-Sybil defense. This necessitates that identity must be anchored to non-transferable biological signatures \cite{ohlhaver2022decentralized}.
    \item \textbf{Inflation-Capital Linkage:} In a hyper-inflationary environment, capital loses the incentive to hold assets, leading to a \textit{bank run} on liquidity pools. Conversely, a stable inflationary environment (through high entropy) encourages capital to flow towards the \textit{labor class} to purchase repair materials or new assets.
\end{itemize}

In conclusion, a failure in any one dimension inevitably compromises the others. An OGE is only as strong as its weakest constraint. This holistic interdependency demands a unified mechanism that addresses all three dimensions simultaneously-a challenge met by the IBAIM model proposed in the following section.

\section{The proposed model IBAIM}

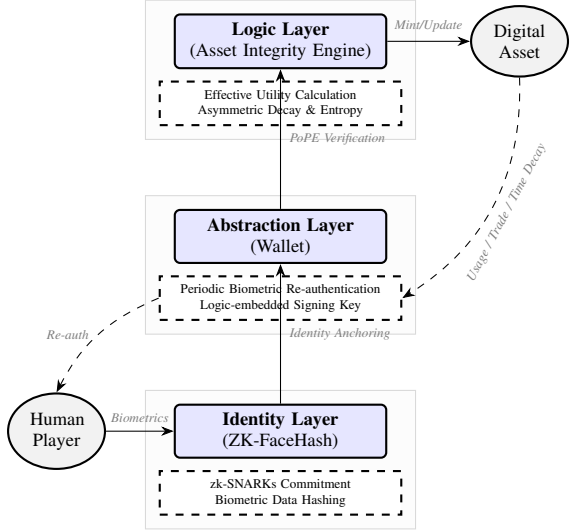
\begin{figure}[!ht]
\centering
\begin{tikzpicture}[
    node distance=1.1cm,
    % 基础样式
    base/.style={draw, thick, align=center, font=\scriptsize, inner sep=3pt},
    % 层级框样式
    layer/.style={base, rectangle, fill=blue!10, text width=2.6cm, rounded corners=2pt, minimum height=0.7cm},
    % 内部组件样式
    comp/.style={base, rectangle, fill=white, text width=3.0cm, dashed, font=\tiny},
    % 外部实体样式
    entity/.style={base, ellipse, fill=gray!10, text width=0.7cm},
    % 箭头样式
    arrow/.style={-Stealth},
    % 注释样式
    tag/.style={font=\tiny\itshape, color=gray}
]

    % --- 1. 逻辑层 (Logic Layer) ---
    \node [layer] (logic) {\textbf{Logic Layer} \\(Asset Integrity Engine)};
    \node [comp, below=0.15cm of logic] (engine) {Effective Utility Calculation\\Asymmetric Decay \& Entropy};

    % --- 2. 抽象层 (Abstraction Layer) ---
    \node [layer, below=1.1cm of engine] (aa) {\textbf{Abstraction Layer}\\(Wallet)};
    \node [comp, below=0.15cm of aa] (auth) {Periodic Biometric Re-authentication\\Logic-embedded Signing Key};

    % --- 3. 身份层 (Identity Layer) ---
    \node [layer, below=1.1cm of auth] (id) {\textbf{Identity Layer} \\(ZK-FaceHash)};
    \node [comp, below=0.15cm of id] (zkp) {zk-SNARKs Commitment\\Biometric Data Hashing};

    % --- 4. 外部实体与流程 ---
    \node [entity, left=0.9cm of id] (player) {Human\\Player};
    \node [entity, right=1.1cm of logic] (asset) {Digital\\Asset};

    % --- 5. 连接线条 ---
    % 生产流程
    \draw [arrow] (player) -- node[above, tag] {Biometrics} (id);
    \draw [arrow] (id) -- node[right, tag] {Identity Anchoring} (aa);
    \draw [arrow] (aa) -- node[right, tag] {PoPE Verification} (logic);
    \draw [arrow] (logic) -- node[above, tag] {Mint/Update} (asset);

    % 生命周期反馈 (衰减与重认证)
    \draw [arrow, dashed, bend left=30] (asset.south) to node[below, tag, sloped] {Usage / Trade / Time Decay} (auth.east);
    \draw [arrow, dashed, bend right=20] (auth.west) to node[left, tag] {Re-auth} (player.north);

    % 背景标注组
    \begin{scope}[on background layer]
        \node [draw, gray!30, fill=gray!2, fit=(logic) (engine), inner sep=5pt] {};
        \node [draw, gray!30, fill=gray!2, fit=(aa) (auth), inner sep=5pt] {};
        \node [draw, gray!30, fill=gray!2, fit=(id) (zkp), inner sep=5pt] {};
    \end{scope}

\end{tikzpicture}
\caption{Technical architecture of IBAIM and the asset integrity life cycle. It demonstrates the vertical integration of biometric identity, account abstraction, and the utility calculation engine.}
\label{fig:ibaim_architecture}
\end{figure}

Building upon the ICR theoretical framework, we propose IBAIM, which is a unified economic architecture designed to enforce labor-value preservation through cryptographic identity anchoring and asymmetric utility scaling.The overall technical architecture of IBAIM, encompassing the vertical integration of biometric identity and asset integrity logic, is presented in Fig. \ref{fig:ibaim_architecture}.

\subsection{Architectural Components}

The IBAIM architecture comprises three primary technical layers:
\begin{enumerate}
    \item \textbf{Identity Layer (ZK-FaceHash):} Utilizing Zero-Knowledge Succinct Non-Interactive Arguments of Knowledge (zk-SNARKs), a player’s biometric data is hashed into a \textit{Commitment} on-chain. This ensures that while the player’s privacy is preserved, their digital output is uniquely tethered to their biological presence.
    \item \textbf{Abstraction Layer:} The system employs Account Abstraction (AA) to decouple the signing key from the account ownership. This allows for complex verification logic-such as periodic biometric re-authentication-to be embedded directly into the wallet’s execution logic.
    \item \textbf{Logic Layer (Asset Integrity Engine):} A set of smart contracts that dynamically calculates the \textit{Effective Utility} of an asset based on its provenance, ownership history, and current durability.
\end{enumerate}

\subsection{Mechanism I: Biometric Tethering and PoPE}

Each asset in IBAIM is minted through a process of PoPE. Upon completion of required in-game activities and time-locks, the asset is minted with a metadata tag containing the $Hash_{Origin}$ of the creator \cite{ohlhaver2022decentralized}.

To ensure an uninterrupted User Experience (UX), Liveness Checks via WebAuthn adopt a 'grace period' strategy, conducting verifications only during natural transition points (e.g., daily logins or before entering high-stakes raids) with advance notifications. If a player fails to authenticate within this given timeframe, only then does the Asset Integrity Engine trigger a utility penalty. This renders \textit{Power Leveling} and bot-farming economically unviable, as the utility is maximized only for the original laborer.

\begin{figure}[htbp]
\centering
\begin{tikzpicture}[
    node distance=1.0cm,
    font=\tiny,
    % Define styles
    layer_box/.style={draw=gray!30, fill=gray!2, dashed, rounded corners, inner sep=6pt},
    comp/.style={draw, thick, fill=white, align=center, minimum height=0.6cm, rounded corners=1pt},
    % Fix: Use a standard rectangle with rounded corners for the shield icon
    shield_style/.style={draw, thick, fill=gray!20, rectangle, rounded corners=3pt, minimum width=0.5cm, minimum height=0.6cm},
    arrow/.style={-Stealth, thick},
    binding/.style={draw=red, ultra thick, {Circle}-{Circle}, shorten >=1pt, shorten <=1pt}
]

    % --- 1. Physical Layer ---
    \node [comp, text width=1.5cm] (phone) {Smart Phone (NFC)};
    \node [comp, left=0.4cm of phone, fill=orange!10, text width=1.0cm] (id) {ePassport /\\National ID};
    \draw [<->, bend left=45, dashed, thick] (id.north) to node[above, font=\fontsize{4}{4}\selectfont] {NFC Scan} (phone.north);
    
    \node [left=0.2cm of id, font=\tiny\bfseries, rotate=90, anchor=south] {Physical};

    \node [comp, above=1.5cm of phone, fill=green!5, text width=1.5cm] (zkpoi) {zk-PoI Generation};
    \node [comp, right=0.3cm of zkpoi, fill=yellow!10, text width=1.5cm] (pseudo) {Pseudo-ID\\($ID_{pseudo}$)};

    \node [shield_style, right=0.3cm of pseudo] (shield) {\checkmark};
    \node [align=center, right=0.01cm of shield, text width=1.8cm, font=\fontsize{5}{6}\selectfont\itshape] {Real ID Data Remains Off-chain};

    \node [left=1.8cm of zkpoi, font=\tiny\bfseries, rotate=90, anchor=south] {Local / Edge};

    \node [comp, above=1.5cm of zkpoi, fill=blue!10, text width=1.6cm, double] (aa) {Wallet\\(Account Abstraction)};
    \node [comp, right=1.2cm of aa, fill=purple!10, text width=1.0cm] (nft) {Game Asset\\(Rare NFT)};
    
    \node [left=1.8cm of aa, font=\tiny\bfseries, rotate=90, anchor=south] {On-chain};

    \draw [arrow] (phone) -- node[right, font=\fontsize{5}{6}\selectfont] {Local Computing} (zkpoi);
    \draw [arrow] (zkpoi) -- (pseudo);
    \draw [arrow] (pseudo) |- node[pos=0.3, right, font=\fontsize{5}{6}\selectfont] {Store $ID_{pseudo}$ only} (aa);

    \draw [binding] (aa.east) -- node[above, color=red, font=\fontsize{5}{6}\selectfont] {HashOrigin} (nft.west);

    % --- 5. Layer Backgrounds ---
    \begin{scope}[on background layer]
        \node [layer_box, fit=(id) (phone)] (box_phy) {};
        \node [layer_box, fit=(zkpoi) (pseudo) (shield)] (box_loc) {};
        \node [layer_box, fit=(aa) (nft)] (box_bc) {};
        
        \node [anchor=south west, gray!80, font=\tiny\bfseries, yshift=2pt] at (box_phy.north west) {Physical Layer};
        \node [anchor=south west, gray!80, font=\tiny\bfseries, yshift=2pt] at (box_loc.north west) {Trusted Local Environment};
        \node [anchor=south west, gray!80, font=\tiny\bfseries, yshift=2pt] at (box_bc.north west) {Blockchain};
    \end{scope}

\end{tikzpicture}
\caption{The privacy-preserving identity anchoring workflow. Biometric verification occurs locally via NFC , while only a pseudo-anonymous identifier is stored on-chain, cryptographically bound to assets.}
\label{fig:privacy_verification}
\end{figure}
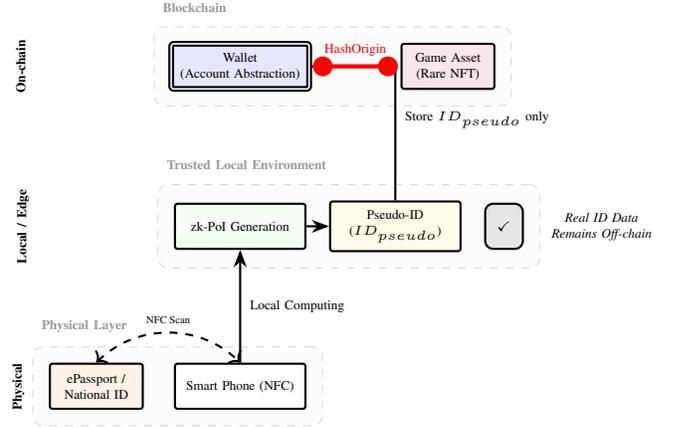

To circumvent the potential compliance and privacy risks associated with the direct collection of biometric data via ZK-FaceHash, the identity verification mechanism can be upgraded through the following measures. First, the system abandons the direct collection and storage of biometric features, instead exogenizing (outsourcing) the identity verification process to government-backed trusted public infrastructures. Second, it leverages the NFC capabilities of modern smartphones combined with trusted hardware to perform local \textit{remote biometric authentication} using widely distributed cryptographic credentials, such as electronic national identity cards and ePassports. Finally, the blockchain exclusively stores a Zero-Knowledge Proof-of-Identity (zk-PoI) alongside a deterministically generated pseudo-anonymous identifier. The user's real-world identity and certificate details remain completely off-chain, thereby achieving absolute privacy while preserving the system's stringent Anti-Sybil resilience.
The privacy-preserving identity anchoring process, which utilizes off-chain biometric verification and on-chain pseudo-anonymous binding, is illustrated in Fig. \ref{fig:privacy_verification}.

\subsection{Mechanism II: Asymmetric Utility and Single-Slot Activation}

To counteract capital-driven dominance, IBAIM introduces  {\bf Asymmetric Utility Decay}. The effective utility $U_{eff}$ of an asset is defined as:
\begin{equation}
    U_{eff} = 
    \begin{cases} 
    U_{base} & \text{if } Hash_{Current} = Hash_{Origin} \\
    0.5 \cdot U_{base} & \text{if } Hash_{Current} \neq Hash_{Origin}
    \end{cases}
\end{equation}

\begin{figure}[htbp]
\centering
\begin{tikzpicture}
    \begin{axis}[
        width=0.48\textwidth, % 适配单栏宽度
        height=5cm,
        xlabel={Time / Usage ($\Delta t, \Delta n$)},
        ylabel={Effective Utility ($U_{eff}$)},
        xmin=0, xmax=10,
        ymin=0, ymax=1.2,
        xtick={0, 5, 10},
        xticklabels={0, $t_{trade}$, $t_{max}$},
        ytick={0, 0.5, 1.0},
        yticklabels={0, 50\%, 100\%},
        legend style={at={(0.5,-0.25)}, anchor=north, legend columns=1, font=\tiny},
        grid=both,
        grid style={line width=.1pt, draw=gray!10},
        major grid style={line width=.2pt, draw=gray!30},
        axis lines=left,
        font=\scriptsize
    ]

    % 1. 传统 Web3 资产曲线 (恒定且永生)
    \addplot [
        domain=0:10, 
        samples=2, 
        color=gray, 
        dashed, 
        thick
    ] {1};
    \addlegendentry{Traditional Web3 Asset (No Decay)}

    % 2. IBAIM 原生持有者曲线 (线性熵增退化)
    \addplot [
        domain=0:10, 
        samples=100, 
        color=blue, 
        thick
    ] {1 - 0.05*x};
    \addlegendentry{IBAIM: Original Creator (Entropy Only)}

    % 3. IBAIM 二手持有者曲线 (包含 50% 不对称衰减断崖)
    % 第一段：交易前
    \addplot [
        domain=0:5, 
        samples=50, 
        color=red, 
        thick,
        forget plot
    ] {1 - 0.05*x};
    
    % 垂直跌落线 (标注交易发生)
    \draw [red, thick, dashed] (axis cs:5, 0.75) -- (axis cs:5, 0.375);
    \node [red, anchor=west, font=\tiny] at (axis cs:5, 0.55) {Secondary Transfer};

    % 第二段：交易后 (效用减半)
    \addplot [
        domain=5:10, 
        samples=50, 
        color=red, 
        thick
    ] {0.5 * (1 - 0.05*x)};
    \addlegendentry{IBAIM: Secondary Holder (Asymmetric + Entropy)}

    % 标注关键点
    \node[anchor=south west, color=blue] at (axis cs:0.5, 1) {$U_{base}$};

    \end{axis}
\end{tikzpicture}

\caption{Comparative analysis of asset utility dynamics. The red curve illustrates the \textit{Asymmetric Utility Decay} (a 50\% drop upon transfer) and the \textit{Entropy-driven Degradation} (continuous slope), which together decouple speculation from in-game merit.}
\label{fig:utility_comparison}
\end{figure}
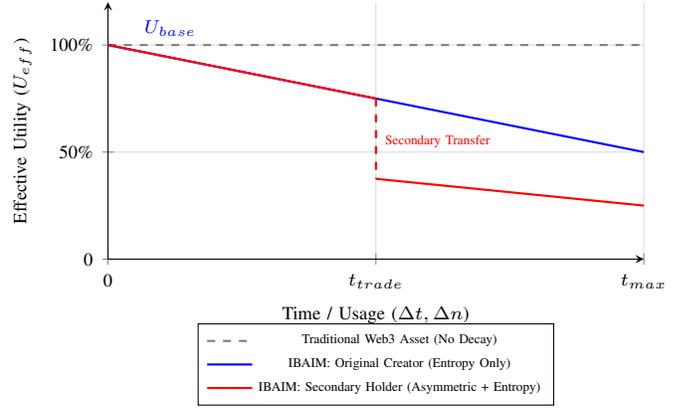

Furthermore, we implement a Single-Slot Activation Rule specifically targeting completely homogenous items: while an account may hold multiple identical high-tier assets, only one such duplicate can be active at any given timestamp $t$. This actively encourages strategic build-crafting, diverse attribute stacking, and set bonuses, while blocking mindless numerical stacking. This constraint ensures that wealth cannot be converted into exponential stat-stacking, maintaining a \textit{Level Playing Field} for merit-based competition.
The dynamic behavior of asset utility under these constraints—specifically the utility cliff during secondary transfer and the subsequent entropy-driven decay—is visualized and compared in Fig. \ref{fig:utility_comparison}.

\subsection{Mechanism III: Dual-Factor Thermodynamic Degradation}

IBAIM treats digital assets as entropic entities. Every asset is subject to a dual-factor durability loss $\Delta D$:
\begin{equation}
    \Delta D = \alpha \cdot \Delta t + \beta \cdot \Delta n
\end{equation}
Where $\Delta t$ represents time elapsed (holding cost) and $\Delta n$ represents usage frequency (activity cost). When $D \to 0$, the asset’s utility $U \to 0$, necessitating repair through the consumption of Repair Materials—resources produced via a PoPE process that specifically rewards players' real skills (e.g., PvP rankings), strategic depth, and social contributions (e.g., guild collaborations and mentoring), effectively transforming mindless grinding into meaningful ecosystem participation.

\subsection{Mapping Analysis: Theory to Implementation}
\begin{table}[htbp] 
\small 
\centering
\caption{IBAIM Mechanism-to-Dimension Mapping}
\label{tab:mapping}
\renewcommand{\arraystretch}{1.3} 
\begin{tabularx}{\columnwidth}{|l|X|l|} 
\hline
\textbf{Mechanism} & \textbf{Economic Logic} & \textbf{Dimension} \\ \hline

ZK-FaceHash & Validates human-to-agent uniqueness; creates infinite marginal cost for bot-farming. & I: Identity \\ \hline

Asymmetric Decay & Imposes a 50\% friction tax on secondary market utility, preserving labor value. & II: Capital \\ \hline

Single-Slot Activation & Prevents utility stacking; caps the ceiling of financial dominance. & II: Capital \\ \hline

Time/Usage Durability & Enforces systemic entropy; maintains continuous demand for new assets. & III: Resource \\ \hline

AA Re-auth. & Cross-Defense: Binds continuous utility to biological signatures to block account trading. & I \& II \\ \hline

\end{tabularx}
\end{table}

Table \ref{tab:mapping} delineates the mapping between the specific technical mechanisms of the IBAIM model and the three fundamental dimensions of the ICR economic framework. It illustrates how ZK-based biometric tethering provides anti-Sybil resilience (Identity), while asymmetric utility decay and single-slot activation collectively mitigate capital dominance (Capital). Furthermore, the integration of usage-based degradation ensures a robust resource sink against inflationary pressure (Resource), with the AA-driven re-authentication acting as a critical cross-defense that reinforces both identity integrity and financial friction.

By integrating these mechanisms, IBAIM ensures that the secondary market acts as a \textit{buffer} for casual players rather than a \textit{ladder} for speculators. The 50\% utility penalty on secondary assets serves as a natural tax on capital, while the degradation ensures that the economy remains a \textit{living} system driven by continuous human participation.

\section{Empirical Analysis: Case Studies}

The validity of the IBAIM model is best demonstrated by analyzing the catastrophic failures of prominent Web3 gaming titles. By examining Axie Infinity, StepN, and CryptoMines, we can observe how the violation of the three fundamental dimensions (Identity, Capital, and Resource) led to their eventual demise and how IBAIM would have provided a systemic defense.

\subsection{Axie Infinity: The Collapse of Identity Integrity}

Axie Infinity pioneered the \textit{Scholarship} model, where \textit{Managers} (capital holders) leased assets to \textit{Scholars} (low-cost laborers). While this initially drove rapid growth, it fundamentally decoupled biological identity from asset production.

\paragraph{Failure Analysis}
The Scholarship model functioned as a decentralized, human-powered Sybil attack. Because there was no biometric tethering, a single biological entity could indirectly control thousands of accounts. This led to an uncontrollable explosion in the minting rate of the Smooth Love Potion (SLP) token, far exceeding the system's absorption capacity.
\paragraph{IBAIM Defensive Value}

Under the IBAIM framework, a direct replication of the Manager-Scholar leasing model would trigger an immediate \textit{ZK-FaceHash} biological mismatch. As the Scholar's biometric signature diverges from the asset's \textit{HashOrigin}, the Asymmetric Utility Decay mechanism forces the leased asset to operate at $0.5 \cdot U_{eff}$. This severe friction tax drastically suppresses the ROI of industrial-scale asset leasing. 

However, to address the more insidious threat of \textit{cheap real-human labor farms}---where a Manager attempts to circumvent the penalty by funding a Scholar to register and mint assets using the Scholar's own authentic biometrics-IBAIM deploys a multi-dimensional defense. First, biometric tethering permanently anchors absolute control of the asset to the Scholar. The Manager loses all cryptographic leverage to seize the asset or enforce revenue sharing, introducing insurmountable trust friction and management risks that destroy the scalability of cross-border, industrialized labor farms. 

Furthermore, the Dual-Factor Thermodynamic Degradation mechanism ensures that repetitive, low-skill grinding accelerates asset decay. Since restoring the asset requires Repair Materials generated exclusively through high-skill, strategic, or socially contributive Proof-of-Play-Effort (PoPE), the maintenance costs of a mechanical labor farm will fatally outpace its yields. Finally, the Single-Slot Activation Rule imposes strict operational and hardware bottlenecks per unique human, completely dismantling the economic viability of capital-driven, exploitative farming models and preserving the ecosystem's value for genuine, authentic players.

\subsection{StepN: Entropy Deficit and Inflationary Saturation}

StepN introduced a \textit{Move-to-Earn} model that utilized repair and minting fees as \textit{sinks.} However, the system eventually succumbed to a supply-side glut.

\paragraph{Failure Analysis}
StepN’s primary failure was in its \textit{Dimension III: Anti-Inflation}. The degradation of assets was static and linear, failing to scale with the total supply of NFT sneakers. As the player base reached its saturation point, the rate of asset minting drastically outpaced the rate of asset destruction. Without a mechanism to enforce systemic entropy, sneakers became a permanent surplus, leading to a \textit{death spiral} in floor prices.
\paragraph{IBAIM Defensive Value}
IBAIM’s \textit{Dual-Factor Thermodynamic Degradation} would have imposed both time-based and usage-based decay ($\Delta D = \alpha \cdot \Delta t + \beta \cdot \Delta n$). By scaling the entropy rate $\alpha$ and $\beta$ relative to total circulating supply, IBAIM would have enforced a mandatory \textit{retirement} of older sneakers, ensuring a perpetual demand for newly minted labor-based sneakers and stabilizing the internal economy.

\subsection{CryptoMines: Frictionless Exit and Capital Predation}

CryptoMines experienced a meteoric rise followed by a total collapse within a 72-hour window, largely due to unchecked capital dominance.

\paragraph{Failure Analysis}
In CryptoMines, capital was perfectly fungible with power. Wealthy actors could instantaneously deploy massive capital to purchase top-tier \textit{Fleets} and extract rewards with zero friction. When the market peaked, these \textit{Whales} exited the system by liquidating their assets instantaneously on secondary markets. The lack of transaction friction or utility penalties on secondary assets allowed for predatory extraction at the expense of long-term stability.
\paragraph{IBAIM Defensive Value}
IBAIM’s \textit{Asymmetric Utility Decay} and \textit{Single-Slot Activation} would have neutralized this threat. A speculator purchasing a \textit{Fleet} on the secondary market would face a 50\% reduction in harvesting efficiency. This \textit{Friction Tax} makes instant capital-driven extraction unprofitable. Furthermore, the single-slot activation would have prevented a single whale from deploying multiple fleets, effectively capping their predatory capacity and allowing the economy to remain meritocratic.

\subsection{Synthesis of Findings}
The synthesized analysis of these historical failures underscores the structural necessity of the IBAIM mechanisms. Whereas Axie Infinity’s reliance on the scholarship model exposed a critical lack of identity integrity—directly addressable by IBAIM’s ZK-FaceHash biometric tethering—StepN’s economic stagnation revealed the fatal consequences of an entropy deficit, which our dual-factor degradation model precisely mitigates by enforcing asset turnover and demand-side renewal. Furthermore, the rapid, capital-led extraction witnessed in CryptoMines demonstrates the systemic peril of frictionless secondary markets, a vulnerability eliminated by the asymmetric utility decay and single-slot activation constraints proposed in this study. Collectively, these empirical findings confirm that the collapses of previous game economies were not anomalous market fluctuations but predictable outcomes of violating the core dimensions of identity, capital, and resource integrity, thereby validating the IBAIM framework as a robust and necessary immune system for sustainable decentralized virtual worlds.

\section{Conclusion}
In conclusion, the sustainability of open game economies is not a byproduct of market sentiment, but a result of rigorous structural design. This paper has demonstrated that Anti-Sybil resilience, Anti-Capital dominance, and Anti-Inflationary saturation constitute the necessary and sufficient conditions for maintaining internal economic equilibrium. Through the proposed IBAIM model, we have provided a technical blueprint that anchors digital utility to biological reality and introduces physical-like entropy through dynamic degradation. While the implementation of asymmetric utility and re-authentication mechanisms necessitates a strategic trade-off in asset liquidity, it represents a vital paradigmatic shift from speculative Ponzi-nomics to a labor-centric value system. Ultimately, the IBAIM framework ensures that the virtual economy remains a meritocratic social space, offering a scalable path toward the long-term viability of decentralized digital civilizations.

Future research will focus on the empirical calibration of the asymmetric decay coefficients and the optimization of Zero-Knowledge proof generation times to further enhance the user experience without compromising systemic integrity.

\section*{Acknowledgment}
%This work reflects the current stage of thinking of the corresponding author (xujinliang@caict.ac.cn; jlxufly@gmail.com), and do not necessarily represent the official views of the affiliated institutions.
The authors contributed collectively to the research. However, the specific strategic visions and conceptual interpretations presented herein are attributed to the corresponding author (xujinliang@caict.ac.cn, jlxufly@gmail.com) to reflect his current stage of thinking, and do not necessarily represent the official views of the other authors or the affiliated institutions.

\bibliographystyle{ieeetr}
\bibliography{Ref}

\end{document}